\documentclass[aps,prx,twocolumn,floatfix,superscriptaddress,footinbib, sort&compress, numbers, merge, reprint]{revtex4-1}

\usepackage{amsmath, amsthm, amsfonts, amssymb}

\usepackage{graphicx} 
\usepackage[dvipsnames]{xcolor} 
\usepackage{verbatim} 

\let\ge\geqslant
\let\le\leqslant

\begin{document}

\title{Functional control of network dynamics using designed Laplacian spectra}

\author{Aden Forrow}
\email{aforrow@mit.edu}
\affiliation{Department of Mathematics, Massachusetts Institute of Technology, 77 Massachusetts Avenue, Cambridge MA 02139-4307, U.S.A.}
\author{Francis G. Woodhouse}
\affiliation{Department of Applied Mathematics and Theoretical Physics, Centre for Mathematical Sciences, University of Cambridge, Wilberforce Road, Cambridge CB3 0WA, U.K.}
\author{J\"orn Dunkel}
\email{dunkel@mit.edu}
\affiliation{Department of Mathematics, Massachusetts Institute of Technology, 77 Massachusetts Avenue, Cambridge MA 02139-4307, U.S.A.}

\begin{abstract}
Complex real-world phenomena across a wide range of scales, from  aviation and internet traffic to signal propagation in electronic and gene regulatory circuits, can be efficiently described through dynamic network models.  In many such systems, the spectrum of the underlying graph Laplacian plays a key role in controlling the matter or information flow.  Spectral graph theory has traditionally prioritized unweighted networks.  Here, we introduce a complementary framework, providing a mathematically rigorous weighted graph construction that exactly realizes any desired spectrum. 
We illustrate the broad applicability of this approach by showing how designer spectra can be used to control the dynamics of various archetypal physical systems. Specifically, we demonstrate that a strategically placed gap induces  chimera states in Kuramoto-type oscillator networks,  completely suppresses pattern formation in a generic Swift-Hohenberg model, and leads to persistent localization in a discrete Gross-Pitaevskii quantum network. Our approach can be generalized to design continuous band gaps  through periodic extensions of finite networks.

\end{abstract}

\maketitle

\section{Introduction}

Spectral band gaps control the behavior of physical systems in areas as diverse as topological insulators~\cite{Hasan2010,Bradlyn2017}, phononic crystals~\cite{Sigmund2003}, superconductors~\cite{Bardeen1957}, acoustic metamaterials~\cite{Wang2017}, and active matter~\cite{Souslov2016}.   In addition to ubiquitous physical network models~\cite{Sonnenschein2013,Bianconi2015,PhysRevLett.117.138301,Tayar2015} ranging from aviation~\cite{Brockmann:2006aa} to electronics~\cite{Cohen:1991aa},
there is also considerable interest in virtual or computational networks~\cite{Yook15102002} with fewer physical constraints, such as those recently used to create spiral wave chimeras in coupled chemical oscillators~\cite{Totz2017}.
Often, dynamics in such systems depend on the graph Laplacian~\cite{McGraw2008,Nakao2010} and in particular on its spectrum of eigenvalues.
Traditionally studied in periodic lattice graph models~\cite{Sigmund2003,Wang2017,Souslov2016,Lubensky2015} and more recently also in hyperuniform systems \cite{Man2013},  the targeted design of spectra of any desired shape remains a major challenge in modern materials science \cite{Wang2017,Han2007}.
Recent breakthroughs in 3D printing~\cite{gladman2016biomimetic,wang2015designable,2016Bhattacharjee_LapChip,huang2017ultrafast} and lithography~\cite{Buckmann2012} make it possible now to produce and explore network structures that go beyond the traditionally considered periodic lattice geometries.

\par
Building on such experimental and theoretical progress, we present here a mathematically rigorous solution to the longstanding question of how any desired spectrum can be realized exactly on a suitably designed positively-weighted network. Specifically, our construction of networks with specified eigenvalues allows us to place arbitrary gaps in the spectrum of the network Laplacian $L = D - A$, where $D$ and $A$ are the weighted degree and adjacency matrices respectively. These gaps, finite analogs to band gaps in continuous systems, enable 
 precise control over the dynamics in a wide range of graph-based physical systems. While in a strict sense band gaps can only exist in an extended system with continuous energy bands, to follow the analogy we will name an eigenvalue-free region in our finite networks that is comparable to the range of eigenvalues a discrete band gap (DBG). That said, our construction can also be used to create continuous band gaps (Section~\ref{s:periodic}). Designing a suitably weighted network topology in this way presents an alternative to control procedures based on adjusting model parameters or initial conditions on a given network~\cite{Motter2015}. The spectral approach towards functional control of network dynamics proposed here can, for example, be directly implemented with recently developed computer-coupled oscillator setups~\cite{Totz2017}.
 
\par
After summarizing the main mathematical result, we demonstrate its broad applicability explicitly for classical and quantum systems, by showing how suitably placed DBGs can induce chimera states~\cite{Martens25062013} in oscillator networks,   inhibit structural growth in generic higher-order pattern formation models, and facilitate state localization in quantum networks. In parallel, we illustrate how our construction can be combined with sparsification algorithms~\cite{Spielman2008,Kelner2013} to yield simplified networks preserving DBGs.   This approach complements the more traditional procedure of constructing graph ensembles with predefined statistical adjacency characteristics~\cite{Barabasi1999, Arenas2006,Peixoto2013, Amir2010}. Finally, we discuss periodic extensions of finite networks as a systematic procedure for designing continuous band gaps.

\begin{figure*}[t]
\includegraphics[width = \linewidth]{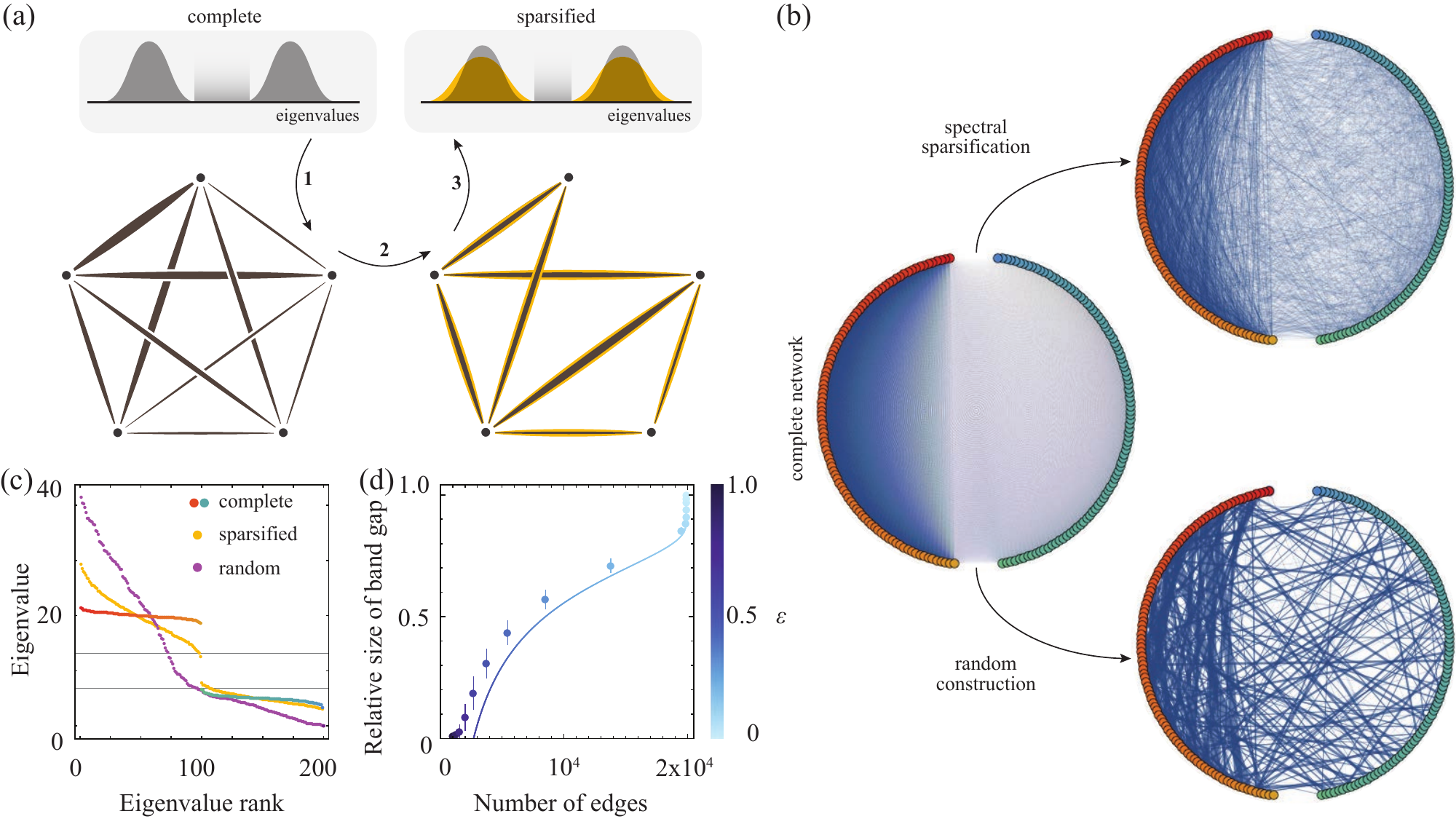}
\caption{ 
Designing networks from spectra.
(a), Schematic of DBG network construction. Given a spectrum of eigenvalues distributed in two (or more) groups, we build a graph with non-negative edge weights that realizes this spectrum exactly~(1). Sparsification of this complete DBG network with the Spielman-Srivastava~\cite{Spielman2008} algorithm (2) yields a new network with wider eigenvalue distributions and a smaller gap (3).
(b), Example graphs used in applications below: Starting from a DBG graph on 200 vertices with 100 eigenvalues set to i.i.d.\ $\mathcal{N}(5, 0.25)$ and 99 set to i.i.d.\ $\mathcal{N}(20,0.25)$ (left), sparsification with $\epsilon = 0.5$ creates a new graph (top) with the number of edges reduced from $19900$ to $3758$.
As a control, we also compare to a gapless random graph (bottom) with  $362$ edges and the same weighted vertex degrees as the original DBG graph (Appendix~\ref{s:random_graph}).
(c), The eigenvalues for the graphs in~(b). The mode on the complete DBG network with the $k$-th largest nonzero eigenvalue is supported on the first $k+1$ vertices, counted counterclockwise from the top red vertex, and highly localized on vertex $k+1$, which is colored to match in (b). Grey lines indicate the borders of the unstable region for the Swift-Hohenberg model with the parameters used in Fig.~\ref{f:SH}.
(d), 
Sparsified networks retain a significant gap even for relatively large $\epsilon$.
Each point shows the mean number of edges and gap size at fixed $\epsilon$ between $1$ (left) and $0.01$ (right), starting from a graph on $200$ vertices designed to have $100 \times$ eigenvalue $5$ and $99 \times$ eigenvalue $20$. The solid curve shows the worst-case gap estimate, reduction by a factor $1-\frac{5}{3}\epsilon$. Sample size is $1000$ for $\epsilon \ge 0.1$ and $300$ for $\epsilon < 0.1$. Error bars are $\pm 1$ standard deviation; horizontal error bars are smaller than the marker size. 
}
\label{f:graphs}
\end{figure*}

\par
\section{Network construction}

The problem of recovering a network from its eigenvalues has been studied extensively, both from an algorithmic~\cite{Ipsen2002, Comellas2008, Cvetkovic2012} and mathematical~\cite{McKay1977,Halbeisen2000} perspective. However, with a few limited exceptions~\cite{Halbeisen2000}, most prior work has focused only on unweighted networks~\cite{VanDam2003}, where there are a finite number of graphs on $n$ vertices and thus only a finite number of possible spectra.
We here construct an exact solution for the weighted case.

Our main result is that, given a set $\{\lambda_i\}$ of desired eigenvalues  ordered so $\lambda_1 \ge \ldots \ge \lambda_{n-1}\ge \lambda_n= 0$, there is a weighted graph $G$ on $n$ vertices with non-negative edge weights whose Laplacian $L$ has spectrum $\lambda_1 ,\ldots, \lambda_{n-1}, 0$. 
The Laplacian, which determines the graph, can be reconstructed from its eigenvalues and eigenvectors with the eigenvalue decomposition; we therefore need to find a set of eigenvectors that together with $\{\lambda_i\}$ give a graph Laplacian.
In fact, the same set of eigenvectors $v^{(k)}, k=1,\ldots, n-1,$ given by
\begin{equation}
v^{(k)}_i = \begin{cases} 
      \frac{1}{\sqrt{k(k+1)}} & i<k+1 \\
      -\frac{k}{\sqrt{k(k+1)}} & i = k+1 \\
     0 & i>k+1
   \end{cases}.
   \label{e:eigenvectors}
\end{equation} 
suffices for any spectrum.
These eigenvectors are strongly localized: the inverse participation ratio (4-norm)
$
\left\| v^{(k)}\right\|_4^4 = 1-2k^{-1} +O(k^{-2})
$ 
indicates near-perfect localization $\left\| v^{(k)}\right\|_4^4 \rightarrow 1$ for almost all $k$, itself a desirable phenomenon \cite{Pradhan2017,Nakao2010}. As the $v^{(k)}$ are mutually orthonormal and orthogonal to the vector of all ones, the matrix  $L = \sum_{k=1}^{n-1} \lambda_k v^{(k)} v^{(k)\top}$ has the desired spectrum with $\frac{1}{\sqrt{n}}{\bf 1}$ as the final eigenvector for $k=n$ with eigenvalue zero. By explicitly computing the sum over $k$ for $i<j$, we find
\begin{equation}
L_{ij}  = \sum_{k=1}^{n-1} \lambda_k u^{(k)}_i u^{(k)}_j
\le -\frac{\lambda_{j-1}}{n},
\end{equation} that is, that the off-diagonal elements of $L$ are all nonpositive (Appendix~\ref{s:positive_edge_weights}); $L$ therefore corresponds to a graph with nonnegative weight $-L_{ij}$ between vertices $i$ and $j$.
If all of the eigenvalues are nonzero, all of the off-diagonal elements of $L$ will be nonzero and the resulting graph will be complete. 

Some spectra can only be realized on complete graphs.
A graph $G$ with approximately constant spectrum must be complete: if $L$ has nonzero eigenvalues $\lambda_k = \lambda + \epsilon_k$ for $k < n$ and $\lambda_n = 0$, then 
\begin{equation}
 \lambda I -L= \frac{\lambda}{n} {\bf 1}{\bf 1}^\top -\sum_{k=1}^{n-1} \epsilon_k v^{(k)}v^{(k)\top}.
\end{equation}
The off-diagonal elements of $ \lambda I -L$, which equal the original edge weights of $G$, are therefore $\frac{\lambda}{n} + O (\epsilon)$. For small $\epsilon_k$, every edge has nonzero weight. More generally, our construction also shows that the spectrum of any non-complete weighted graph with no isolated vertices cannot uniquely specify that graph, in line with older results on, for example, the spectra of trees~\cite{McKay1977}.

\begin{figure*}
\includegraphics[width = \linewidth]{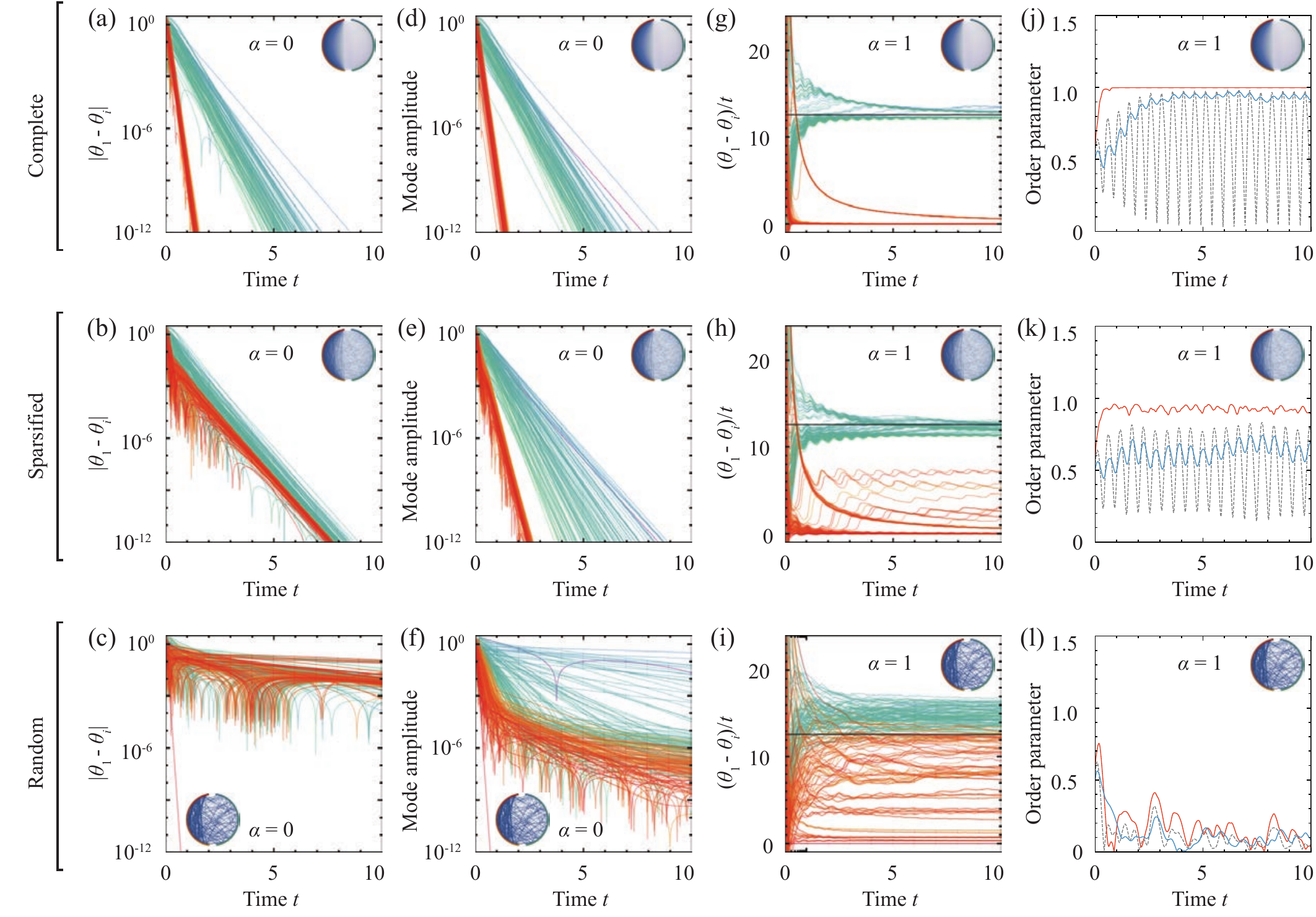}
\caption{DBG networks lead to staggered synchronization and chimeras. 
(a-f), In the Kuramoto model with $\alpha = 0$, the complete (first row) and sparsified (second row) graphs synchronize much faster than the random graph (third row). For the complete graph the gap affects the rate of synchronization, with highly-connected vertices synchronizing faster~(a), while on the sparsified graph the gap is only visible in the mode basis~(e).
(g-i), Chimera states appear when $\alpha = 1$. Both the complete (g) and sparsified (h) graphs have two dominant groups of phase-locked oscillators, with the complete graph more fully synchronized. Dynamics on the random graph (i) are much less coherent.
Solid black lines indicate the predicted approximate frequency difference for a network with two distinct eigenvalues, 5 and 20.
(j-l), Order parameter $r = | \sum_j e^{i \theta_j} |$ for the simulations in (g-i) for the strongly-connected vertices (red), weakly-connected vertices (teal), and all vertices (gray). See Movie 1 for animation.
}
\label{f:Kuramoto}
\end{figure*}

\par
This construction allows us to create networks with precisely specified gaps. For instance, choosing $\lambda_1 = \lambda_2 = \ldots = \lambda_{n/2-1}$ and $\lambda_{n/2} = \lambda_{(n/2) +1} = \ldots = \lambda_{n-1}$ leads to a graph with edge weights $-L_{ij} = \lambda_{n-1}/n$ if $i>n/2$ or $j>n/2$ and $-L_{ij} = (2\lambda_1 - \lambda_{n-1})/n$ otherwise (Appendix~\ref{s:gap_weights});
that is, there are two groups of vertices, one strongly connected within itself and one weakly connected to everything. Adding a small amount of noise to each eigenvalue then lifts the eigenvalue degeneracy while preserving the connectivity structure and retaining a gap (Fig.~\ref{f:graphs}b,c).

\par
Since complete graphs can be difficult to realize physically, we explore the effect of the sparsification-by-resistances algorithm developed by Spielman and Srivastava \cite{Spielman2008}. Given an accuracy parameter $\epsilon$, this sparsification creates a network with $O(n \log(n)/\epsilon^2)$ edges whose eigenvalues match the eigenvalues of the original network to within a multiplicative factor $1\pm \epsilon$ with high probability.
Sparsification by resistances aims to preserve the entire spectrum, not just a gap; future sparsification algorithms directly constructed to preserve a gap could therefore improve on its efficiency. In other applications, the networks of interest are virtual ones~\cite{Totz2017}
and sparsification may not be necessary.

We can use the $1\pm \epsilon$ multiplicative error bound to estimate the size of a discrete band gap after sparsification. Suppose we start from a network with eigenvalues $\lambda_1$,  $\lambda_2$, and $0$, with some multiplicities, where $\lambda_1 > \lambda_2$. The eigenvalues $\{\mu_i\}$ of the sparsified graph corresponding to $\lambda_1$ should be no smaller than $\mu_i \ge (1 - \epsilon ) \lambda_1$, while the eigenvalues $\{\nu_i\}$ corresponding to $\lambda_2$ should be no larger than $\nu_i \le (1 + \epsilon) \lambda_2$. The sparsified graph should therefore have a gap $\Delta = \min_i \mu_i - \max_i \nu_i$ of size
\begin{align}
\Delta &\ge \left( 1 - \frac{\lambda_1 + \lambda_2}{\lambda_1 - \lambda_2} \epsilon \right) (\lambda_1 - \lambda_2).
\end{align}
That is, the gap contracts by a factor at most $1 - \frac{\lambda_1 + \lambda_2}{\lambda_1 - \lambda_2} \epsilon$. For the parameters used in Fig.~\ref{f:graphs}d, this is $(1 - \frac{5}{3}\epsilon)$.

\par
\section{Applications}
We now demonstrate the practical potential of DBGs with three generic network models. In each case, we compare the dynamics on a complete DBG network (Fig.~\ref{f:graphs}b, left) both to a sparsified approximate DBG network (Fig.~\ref{f:graphs}b, top) and to a random connected network (Fig.~\ref{f:graphs}b, bottom) constructed to have the same weighted vertex degrees as the DBG network (Appendix~\ref{s:random_graph}). The gap is approximately preserved in the sparsified network  and vanishes entirely in the random graph (Fig.~\ref{f:graphs}c,d). 
Matching the degrees in the random graph to the DBG network ensures that any differences in dynamics are due to the gap and not differences in coarse features like the average connectivity. Often, the behavior of optimized networks is sensitive to small perturbations~\cite{Nishikawa2017}; here, behaviors preserved in the sparsified graph are robust to significant changes.

Simulations were performed using a third or fourth order Adams-Bashforth linear multistep method with a time step $\Delta t = 10^{-4}$. All simulations were written in C++ using Armadillo \cite{Sanderson2016}.

\begin{figure*}
\includegraphics[width = \linewidth]{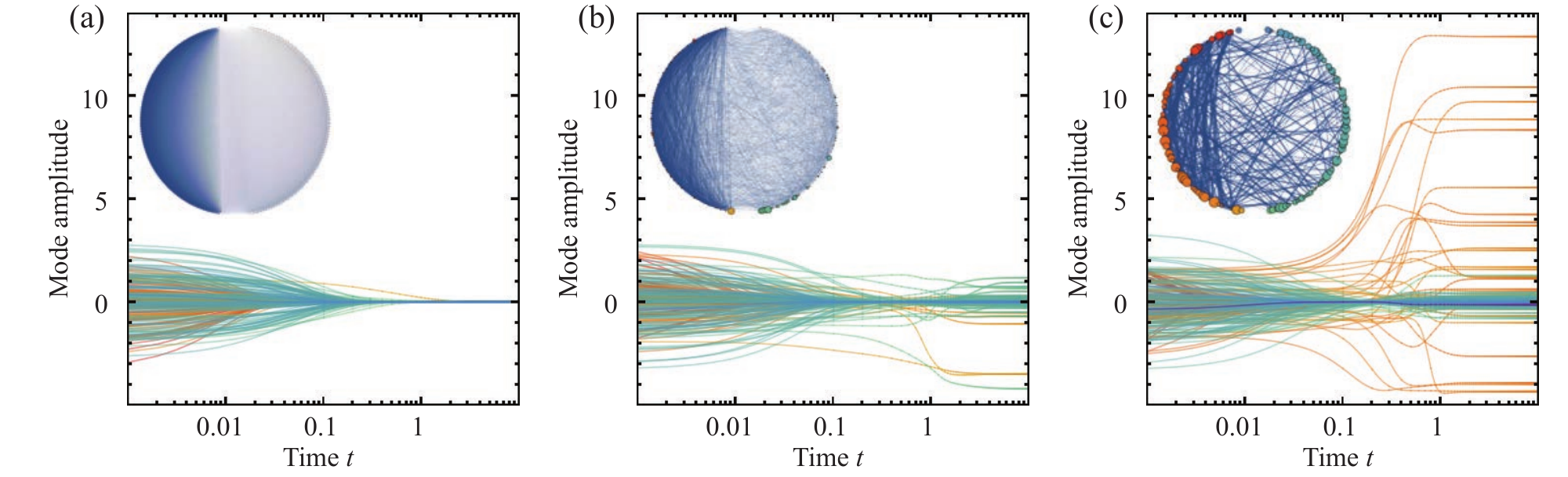}
\caption{ Generic suppression of pattern formation with a designed discrete band gap. 
(a),~Pattern formation in the Swift-Hohenberg system is completely suppressed by constructing a gap around the range of unstable eigenvalues (Fig.~\ref{f:graphs}c). 
(b),~On a sparsified graph that has a few eigenvalues just within the unstable region, some modes settle at small nonzero values. 
(c),~On the random graph many more eigenvalues are well within the unstable region and the corresponding modes settle at larger amplitudes.
Inset graphs show the final steady state on each graph.
All simulations used identical initial conditions $u_i \sim \mathcal{N} ( 0, 1)$ and parameters $\alpha = 90$, $D_1=-20$,  $D_2 = 1$. See Movie 2 for animation.}
\label{f:SH}
\end{figure*}

\par
\subsection{Kuramoto oscillators}
Our first application is the Kuramoto model of coupled oscillators \cite{Kuramoto1975,Strogatz2000}. Recent experiments coupling Belousov-Zhabotinsky reactions via a computer-controlled projector have shown the emergence of chimeras~\cite{Totz2017}; we will show this can be achieved in the Kuramoto model by designing an appropriately gapped spectrum. Here phases $\theta_i$ on the vertices evolve with a natural frequency $\omega$ and a nonlinear coupling defined by the network adjacency matrix:
\begin{equation}
\frac{d\theta_i}{dt} = \omega + \sum_{j=1}^n A_{ij} \sin(\theta_j - \theta_i + \alpha).
\end{equation}
On any connected graph with $\alpha = 0$, there is a single global attractor $\theta_i =\theta_0+ \omega t $. The rate of convergence to this attractor is controlled by the eigenvalues of the Laplacian $L$~\cite{McGraw2008}. Both the complete and sparsified graphs have no eigenvalues near zero, so they synchronize much faster than the random graph (Fig.~\ref{f:Kuramoto}a-c). The gap divides the modes into two groups, one synchronizing faster than the other (Fig.~\ref{f:Kuramoto}d-e); moreover, on the complete graph, the localization of the eigenvectors causes staggered synchronization of vertices (Fig.~\ref{f:Kuramoto}a).

\par
If $\alpha$ is sufficiently large, the oscillators no longer synchronize at a single frequency. On DBG networks, global coherence gives way to weak chimera states \cite{Ashwin2015} where vertices synchronize into two clusters with distinct frequencies (Fig.~\ref{f:Kuramoto}g-i, Movie 1). 
For the exactly-gapped network with edges of weight $w_1 = \lambda_{n-1}/n + (\lambda_1 - \lambda_{n-1})/(m+1)$ or $w_2 = \lambda_{n-1}/n$ there is a steady state with $\theta_i = \theta_1$ for $i \le m+1$ and $\theta_i = \theta_n$ for $i > m+1$. In this state,
\begin{align}
\frac{d}{dt} (\theta_n - \theta_1) & = \left[ nw_2 - (m+1)(w_1+w_2) \right] \sin(\alpha) \nonumber
\\
& \qquad - n w_2 \sin(\theta_n - \theta_1 + \alpha).
\end{align}
The two phases $\theta_1$ and $\theta_n$ can synchronize if
\begin{align}
\sin(\theta_n - \theta_1 + \alpha)  
& =  \left[ 1 -2 \frac{m+1}{n} - \left(\frac{\lambda_1}{\lambda_{n-1}} - 1\right) \right] \sin(\alpha).
\label{e:phase_difference}
\end{align}
This synchronization is possible if $\alpha$ is small enough that the right hand side is less than one. If the two groups do not synchronize, and $nw_2 = \lambda_2$ is not too large, the sine term in Eq.~\eqref{e:phase_difference} will average to nearly zero giving an approximate mean frequency difference
\begin{align}
\biggl\langle \frac{d}{dt} (\theta_n - \theta_1) \biggr\rangle \approx \left[ nw_2 - (m+1)(w_1+w_2) \right] \sin(\alpha) ,
\end{align}
which reduces to $-(\lambda_1 - \lambda_{n-1}) \sin(\alpha)$ if $m+1 = \frac{n}{2}$ as in Fig.~\ref{f:Kuramoto}. More general cluster synchronization~\cite{Cho2017,Pecora2014} could be achieved by adjusting the number and size of the gaps.
In contrast, the random graph becomes thoroughly incoherent at comparable values of $\alpha$ (Fig.~\ref{f:Kuramoto}i,l). The coherence can be quantified by the order parameter $r = | \sum_j e^{i \theta_j} |$, which oscillates for the complete and sparsified networks but is near zero for the random graph (Fig.~\ref{f:Kuramoto}j-l), indicating complete disorder.

\par
\subsection{Swift-Hohenberg pattern formation}
As the second application,  we study generic Swift-Hohenberg pattern formation dynamics on a network~\cite{Swift1977,Nicolaides2015}. Consider a scalar field $u_i$ on the vertices obeying
\begin{equation}
\frac{d u_i}{dt} = -D_1 \sum_{j=1}^n L_{ij}  u_j - D_2 \sum_{j,k = 1}^n L_{ij}L_{jk} u_k - \alpha u_i -u_i^3.
\end{equation}
The fixed point $u_i = 0$, which exists for any values of the parameters $D_1$, $D_2$ and $\alpha$, is linearly stable to perturbations in a Laplacian eigenmode with eigenvalue $\lambda$ if the growth rate $\sigma\equiv -\alpha - D_1 \lambda - D_2 \lambda^2 < 0$. 
With $\alpha$ and $D_2$ positive, $\sigma$ is negative for small and large $\lambda$, but choosing $D_1 < -2\sqrt{\alpha D_2}$ creates a range of unstable~$\lambda$ in between. This can drive pattern formation that is eventually stabilized by the nonlinearity. The patterns can only form, however, if $L$ has eigenvalues in the unstable range. Controlling the spectrum of $L$ therefore allows us to completely suppress pattern formation in arbitrarily large systems by placing a  gap around the unstable region (Figs.~\ref{f:graphs}c,~\ref{f:SH}a, Movie 2). If we sparsify the network with sufficiently small $\epsilon$, the gap will be preserved and again no patterns will form. Eventually, though, increased sparsification will push some eigenvectors into the edges of the unstable region and bring back partial pattern formation (Fig.~\ref{f:SH}b), which becomes fully developed in the random graph (Fig.~\ref{f:SH}c). The maximum $\epsilon$ for which patterns will be fully suppressed for given parameter settings can be predicted straightforwardly from the expected changes in the eigenvalues, in a similar fashion to the post-sparsification gap size in Fig.~\ref{f:graphs}d.

\begin{figure}[b]
\center
\includegraphics[width =\linewidth]{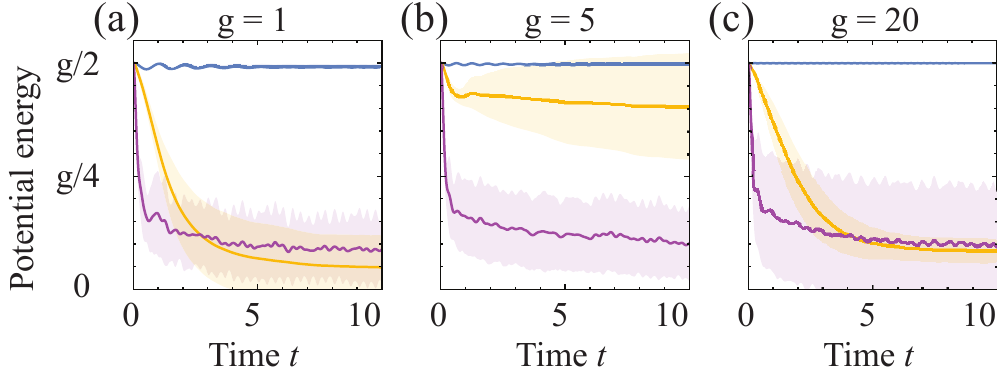}
\caption{Localization on a DBG quantum network (Movie 3).
\mbox{(a-c),}~When the wavefunction in the Gross-Pitaevskii model of Eq.~\eqref{e:GP} is initialized at a weakly connected vertex with low kinetic energy, localization or delocalization (indicated by high or low potential energy, respectively) is controlled by the interplay between the graph spectrum and the rate of potential energy loss $g$.
The random graph (purple) always delocalizes, due to its dense spectrum.
However, while the sparsified graph (yellow) can delocalize for low $g$ (a) and high~$g$~(c), again due to available eigenmodes, intermediate $g$ (b) places the range of allowed modes inside the spectral gap, preventing delocalization.
The complete graph (blue) always inhibits spreading due to the extreme localization of its eigenvectors.
}
\label{f:GP}
\end{figure}

\subsection{Gross-Pitaevskii localization}
Having discussed two classical applications to non-conservative systems, we now show how DBGs can control quantum dynamics with conserved energy.
In a network version of the Gross-Pitaevskii model
~\cite{Brunelli2004,Bang1994,Pelinovsky2014}, similar to those used in studying Bose-Einstein condensates in optical lattices~\cite{Greiner2002}, 
 we find that the interplay of the total energy conservation constraint in such a model with  the kinetic energy gap inhibits spreading of the wavefunction on DBG networks. Similarly to the Swift-Hohenberg example, we take the Gross-Pitaevskii equation for a  complex wavefunction $\psi$ and replace the continuous Laplacian $\nabla^2$  with its discrete analog $-L$:
\begin{equation}
i \frac{d\psi_j}{dt} = \sum_{k=1}^n L_{jk}\psi_k + g |\psi_j|^2  \psi_j.
\label{e:GP}
\end{equation}
This can be written $i \frac{d\psi_i}{dt} = \frac{\partial E}{\partial \psi^*_i}$, where the energy $E$ is the sum of the kinetic energy $T = \sum_{i,j} \psi_i^* L_{ij} \psi_j$ and the potential energy $V =\frac{1}{2} g  \sum_j (\psi^*_j \psi_j)^2$. 
The potential energy quantifies the localization of $\psi$: with $g>0$, it is large when the probability $\psi^*\psi$ is concentrated at a single vertex and small when $\psi^*\psi$ is spread out. Delocalization is limited by the size of the network, as $V \ge \frac{g}{2n}$, but can vary widely even on a finite network.
If $\psi$ is initialized at a single vertex $j$, then $V = g/2$, independent of $j$, while $T = L_{jj}$ equals the degree of $j$. 

Since energy is conserved, the wavefunction can delocalize and reduce its potential energy only by converting it to kinetic energy. The rate of potential energy loss, set by $g$, must therefore match the rate of kinetic energy gain, set by the differences in eigenvalues among the modes involved.
Suppose the wavefunction is mostly in a localized mode $j$ with eigenvalue $\lambda_j$. Spreading to a higher mode $k$ with $\lambda_k - \lambda_j \gg g$ would increase kinetic energy by more than it would decrease potential energy, while a weak higher mode $0 < \lambda_k - \lambda_j \ll g$ or a lower mode $\lambda_k < \lambda_j$ would not increase kinetic energy by enough, if at all.
Both are barred by energy conservation. The amplitude in mode $j$ can only be reduced if there are other modes $k$ with $\lambda_k \sim \lambda_j + g$.

\begin{figure*}[t]
\includegraphics[width = \linewidth]{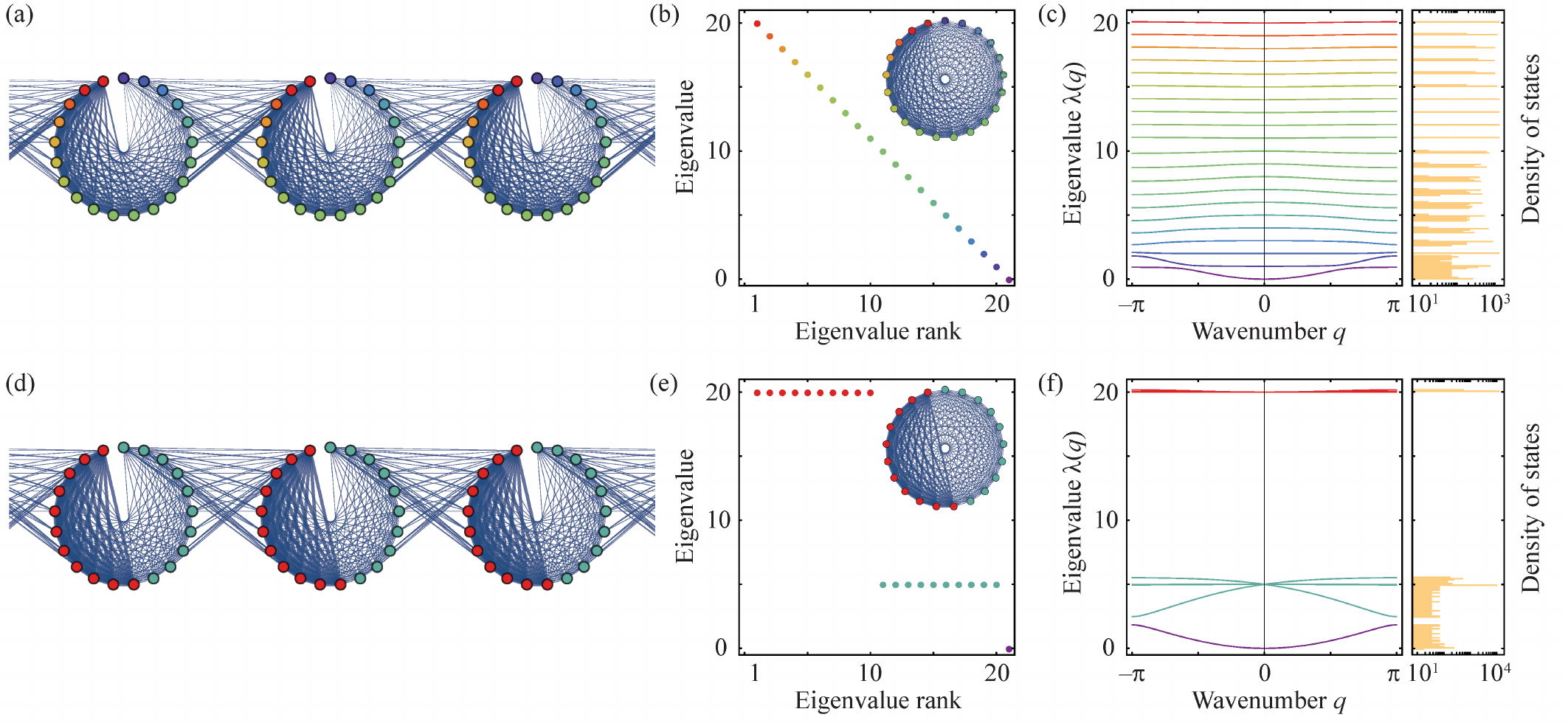}
\caption{ 
Designed spectra on a discrete network are preserved when extended periodically in one dimension.
(a)
We extend a finite network to an infinite one
by  rewiring a subset of the edges to cross between adjacent copies of the original network. Here, we take the network with the spectrum in (b) and rewired the edge between vertices $j$ and $k$ if $|k-j| > n/2$. This rewires roughly one quarter of the edges.
(b) One unit cell in (a) would have a discrete spectrum with $\lambda_j = 21-j$.
(c) Most of the eigenvalue bands do not change significantly with $q$, so the density of states consists of 21 sharp peaks with low- or zero-density regions between.
(d) The same construction as in (a) can be repeated for any spectrum; this is the result for a gapped network.
(e) One unit cell in (d) would have a gapped spectrum, with 10 eigenvalues equal to 20 and 10 equal to 5, in addition to the always-present zero eigenvalue.
(f) Again, most of the eigenvalue bands are roughly constant, even though the eigenvectors do depend strongly on $q$. The gap in the middle of the spectrum is nearly perfectly preserved; a small gap remains between the bottom two bands. Note the log scale on both density of states plots.
}
\label{f:periodic}
\end{figure*}

To see this in more detail, suppose we have a wavefunction comprising two modes, $\psi_j = c_1 v^{(1)}_j + c_2 v^{(2)}_j$, with initial complex amplitudes $c_1$, $c_2$. 
Suppose also that these eigenmodes are localized on two different vertices, with $v^{(1)} \approx (-1, \epsilon, \epsilon, \ldots, \epsilon)$ and $v^{(2)} \approx (\epsilon, -1, \epsilon, \ldots, \epsilon)$.
The system energy as a function of $c_1$ and $c_2$ is then
\begin{align}
E = \lambda_1 |c_1|^2 + \lambda_2 |c_2|^2 + \tfrac{1}{2}g\left[ |c_1|^4 + |c_2|^4 + O(\epsilon) \right].
\end{align}
If the squared amplitudes change slightly, to $|c_1|^2 - \delta$ and $|c_2|^2 + \delta$, the change in energy to leading order in $\delta$ is
\begin{align}
\Delta E =\left[ \lambda_2 - \lambda_1 + g \left( |c_2|^2  -   |c_1|^2 \right) \right] \delta + O(\epsilon).
\end{align}
Conservation of energy requires $\Delta E = 0$, so in order to transfer a noticeable amplitude $\delta \gg \epsilon$ from the first mode to the second we must have $\lambda_2 - \lambda_1 + g(|c_2|^2  -   |c_1|^2) \approx 0$. In the cases considered in Fig.~\ref{f:GP}, where $|c_1| \approx 1$ and $|c_2| \approx 0$, this reduces to $\lambda_2 - \lambda_1 \approx g$.
 Thus on a network with a spectral gap, the localization of $\psi$ can depend non-trivially on the interplay between $g$ and the spectrum.

Initializing $\psi$ at a weakly-connected vertex brings out this interplay as~$g$ is varied (Movie 3).
The initial state, with high potential energy and low kinetic energy, is localized on modes with eigenvalue below the spectral gap.
On the sparsified network, a low value of~$g$ makes nearby modes below the gap accessible for delocalization, causing the wavefunction to spread (Fig.~\ref{f:GP}a).
However, increasing~$g$ pushes the region where transfer is possible inside the spectral gap, inhibiting the spread of the wavefunction on the sparsified network (Fig.~\ref{f:GP}b).
Further increase of~$g$ once again enables delocalization as the modes above the gap becomes accessible for energy transfer (Fig.~\ref{f:GP}c).
In contrast, the dense spectrum of the random graph means delocalization occurs in all three instances (Fig.~\ref{f:GP}).
Interestingly, the complete DBG network appears to remain localized for all values of~$g$ in our simulations (Fig.~\ref{f:GP}); this is likely due to the strong localization and near-zero overlap of the eigenmodes.

\section{Band structure in periodic networks}
\label{s:periodic}
We can construct infinite periodic networks in a standard way from any base network $G$ by tiling periodically and rewiring edges (Fig.~\ref{f:periodic}a,d). Starting from the original vertex set $\{j\}$ for $1 \le j \le n$ and edge weights $-L_{jk}$, we make an infinite string $G^\infty$ of copies of $G$ with vertices indexed by $j$, the label in $G$, and $c \in \mathbb{Z}$, the unit cell. This will give a new, infinite Laplacian $L^{\infty}$. For the edges that will not be rewired, we set $L^\infty_{jc, kc} = L_{jk}$ for all $c$. Doing this for all edges would leave the copies of $G$ disconnected. To connect them, we choose a subset of edges $\{(j,k)\}$ and rewire them to cross between unit cells; for example, if $(j,k)$ is an edge to be rewired to have $k$ in a unit cell to the left of $j$ we can set $L^\infty_{jc, k(c-1)} = L_{jk}$ for all $c$ and symmetrically set $L^\infty_{k(c-1), jc} = L_{kj}$.
The remainder of the entries of $L^{\infty}$ are set to zero.

Since $L^\infty$ is periodic, Bloch's Theorem allows us to write the eigenvectors as
\begin{equation}
U^{\infty}_{jc}(q) = e^{i q c} \tilde U_{j}(q),
\end{equation}
where $q$ is a wavenumber in the first Brilloun zone $-\pi < q < \pi$.
The $\tilde U$ then satisfy
\begin{equation}
\lambda(q) e^{i q c} \tilde U_{j} (q)= \sum_{k,d} L^{\infty}_{jc,kd} e^{i q d} \tilde U_{k}(q),
\end{equation}
which reduces to a new eigenvalue equation for a matrix of size $n$:
\begin{equation}
\lambda(q)  \tilde U(q) = \tilde L(q)  \tilde U(q),
\end{equation}
where the matrix elements of $\tilde L(q)$ are the same as those of $L$ for edges within a single unit cell and differ by a factor $e^{i q(c -  d)}$ for edges that cross between unit cells $c$ and $d$.

Using these transformations, which are standard in the study of lattice systems~\cite{Lubensky2015}, we can find the continuous spectra of periodic tilings of our designed networks. Even without any optimization of which edges to rewire, the spectral characteristics persist in the infinite system. If we rewire all edges with $|j-k| > n/2$, for example, a spectrum of equally-spaced eigenvalues leads to a density of states with corresponding equally-spaced large spikes (Fig.~\ref{f:periodic}a-c), while a discrete band gap is almost entirely preserved (Fig.~\ref{f:periodic}d-f).
In both cases, only the bottom few bands vary significantly with $q$. Note that, because we moved edges incident to the first vertex, all of the eigenvectors do change and are not localized for nonzero~$q$.

\par
\section{Conclusions}
Controlling dynamics on a network typically requires detailed understanding of its spectral properties. Here we have reversed the conventional approach by starting from a desired spectrum and providing a mathematically rigorous construction of a matching network. This enabled us to induce chimera states, suppress pattern formation, and control wavefunction localization~\cite{Amir2013} using suitably designed gapped spectra. 
Our method, which starts from global properties, complements traditional approaches using small-scale local rules to build and analyze networks \cite{Barabasi1999,Bianconi2001,Lorimer2015,Milo2002}. In the future, the above results may also prove useful as a standard of comparison for other networks. Contrasting the dynamics on an important class of networks with the dynamics on networks designed to have identical spectra can help identify the important features of that class. Moreover, as dynamics are often related to matrices other than the Laplacian~\cite{Yan2015}, it will be interesting to investigate control of their spectra for weighted networks as well. 
Although our construction works optimally with fully-connected graphs, one can expect that improved sparsification algorithms together with recent progress in 3D printing and lithography~\cite{Buckmann2012, Man2013}
may soon lead to physically-realizable networks with arbitrary gaps; since any graph can be embedded in 3D~\cite{Cohen1997}, the framework introduced here lays a conceptual foundation for the targeted design of complex non-periodic metamaterials with desired spectral properties. Currently, the approach can be applied directly to interfacing biochemical oscillators with computational networks~\cite{Totz2017}.  Such hybrid networks can also be naturally extended to periodic systems, where the spectral properties are preserved well without any further optimization.

\par
\textbf{Acknowledgements.}
The authors would like to thank Jon Kelner and Philippe Rigollet for helpful discussions. This work was supported by Trinity College, Cambridge (F.G.W.),  an Edmund F. Kelly Research Award (J.D.), and a Complex Systems Scholar Award of the James S. McDonnell Foundation (J.D.).

\appendix

\section{Edge weights with a gap}
\label{s:gap_weights}
Suppose we have eigenvalues $\lambda_1$ with multiplicity $m$ and $\lambda_{n-1} < \lambda_1$ with multiplicity $n-m-1$. Then if $i < j \le m+1$
\begin{align}
L_{ij} &=  - \frac{ \lambda_1}{j} + \sum_{k=j}^{m}\frac{\lambda_1}{k(k+1)}+ \sum_{k=m+1}^{n-1}\frac{\lambda_{n-1}}{k(k+1)}\nonumber
\\
& =  - \frac{ \lambda_1}{m+1} + \lambda_{n-1}\left( \frac{1}{m+1} - \frac{1}{n}\right).
\end{align}
Else, if $i<j$ and $j > m+1$,
\begin{align}
L_{ij} &=\lambda_{n-1}\left[  - \frac{ 1}{j} + \sum_{k=j}^{n-1}\frac{1}{k(k+1)}\right]
=- \frac{\lambda_{n-1}}{n}.
\end{align}
There are two types of edges: edges with both endpoints in the first $m+1$ vertices have weight $\lambda_{n-1}/n + (\lambda_1 - \lambda_{n-1})/(m+1)$, while other edges have weight $\lambda_{n-1}/n$.

\section{Positivity of edge weights} 
\label{s:positive_edge_weights}
The elements of the designed $L$ above the diagonal, $L_{ij}$ for $i<j$, are given by
\begin{align}
L_{ij} & = \sum_{k=1}^{n-1} \lambda_k u^{(k)}_i u^{(k)}_j\nonumber
\\
& = \lambda_{j-1} u^{(j-1)}_i u^{(j-1)}_j + \sum_{k=j}^{n-1} \lambda_k u^{(k)}_i u^{(k)}_j\nonumber
\\
& \le \lambda_{j-1}\left[ -\frac{1}{j} + \sum_{k=j}^{n-1} \frac{1}{k(k+1)} \right]\nonumber
\\
& = \lambda_{j-1}\left( -\frac{1}{j} + \frac{1}{j} - \frac{1}{n}\right)\nonumber
\\
& = -\frac{\lambda_{j-1}}{n} \le 0.
\end{align}
From the second to third lines we use the definition of the eigenvectors in Eq.~\eqref{e:eigenvectors}; the sum $\sum_{k=j}^{n-1} \frac{1}{k(k+1)} = \frac{1}{j}-\frac{1}{n}$ in the third line can be computed as a telescoping sum of partial fractions. 
$L$ is symmetric, so the elements below the diagonal must also be nonpositive. This proves that the edge weights of the constructed graph are nonnegative.

\section{Random equal-degree graphs}
\label{s:random_graph}
Given a weighted graph $G$, we can construct a random simple graph $\tilde G$ with the same vertex degrees as $G$ as follows. Let $w(e)$ denote the weight of edge $e$ and $d(v)$ denote the weighted degree of vertex $v$. Begin with a disconnected graph with a loop of weight $d(v)/2$ at each vertex $v$; this has the same degrees as $G$ but is not simple. Repeat the following steps until there are no loops:
\begin{enumerate}
\setlength{\itemsep}{1pt}
\item
Pick a loop $l = (u,u)$ and another edge $e = (v,w)$ at random, with $v\neq u \neq w$. 

\item
\begin{enumerate}
\setlength{\itemsep}{1pt}
\item If $w(l) > w(e)$, remove $e$ and add $e' = (u,v)$ and $e'' = (u, w)$ with weight $w(e)$. Subtract $w(e)$ from the weight of $l$.
\item Else, remove $l$ and add $e' = (u,v)$ and $e'' = (u, w)$ with weight $w(l)$. Subtract $w(l)$ from the weight of $e$.
\end{enumerate}
\end{enumerate}
Once there are no more loops, merge all sets of edges between the same pair of vertices into one edge with the same total weight. Since the degree of each vertex is preserved at each step, the final graph has the same degrees as $G$. In the examples considered here, the algorithm terminates quickly.


%

\end{document}